\documentclass[12pt]{iopart}
\usepackage{graphicx}
\usepackage{iopams}

\begin{document}

\title[A THz time-domain susceptibility for superconductors including...]{A THz time-domain susceptibility for superconductors including strong-current effects}

\author{Xiaoxiang Xi and G L Carr}

\address{Photon Sciences, Brookhaven National Laboratory, Upton, NY 11973, USA}
\ead{carr@bnl.gov}

\begin{abstract}
Finite-difference time-domain methods are increasingly being used to develop, model and analyze the response of materials, including engineered metamaterials that may contain superconductors. Though simple and useful expressions for the time-domain susceptibility exist for basic metals and dielectrics, the time-domain response for a superconductor has not been developed, mainly because the frequency-dependent expressions themselves are rather complex. In this paper we present a simple approximate expression for the time-domain susceptibility of a superconductor for the $\hbar/2\Delta$ time scale (where $\Delta$ is the BCS energy gap) that fulfills causality requirements, and demonstrate its ability to model the transmission and reflection of a fully-gapped superconductor in the THz region. By allowing $\Delta$ to be a function of current, we also show how this model function can be used to describe nonlinear microwave response in superconductors.
\end{abstract}

\maketitle

\section{Introduction}
The use of time-domain analysis of electromagnetic problems has increased with the recent focus on plasmonics and engineered metamaterials for tailored optical properties \cite{Lal2007}. Algorithms based on the finite-difference time-domain (FDTD) approach to electromagnetic wave propagation are now in common use for spectral ranges extending up through the THz and infrared \cite{Semouchkina2005}. In the time domain, the fields and currents that result in response to an applied E-field $E(t)$ are given by a convolution of $E(t)$ with the appropriate response function: the susceptibility $\chi(t)$ for determining the polarization $P(t)$ and the conductivity $\sigma(t)$ for determining the current density $J(t)$. If one removes the artificial distinction between the currents associated with free charge and those for bound charge, then $J(t)$ = $\partial P(t)/\partial t$ and the conductivity $\sigma(t)$ is the time-derivative of the susceptibility $\chi(t)$, i.e. $\sigma(t) = \partial \chi(t)/\partial t$. A benefit of the time domain approach is the ability to incorporate a non-linear response determined by the instantaneous strength of the induced current density.

The electrodynamic response for many common materials can be described adequately using classical Lorentzian oscillators having explicit algebraic forms in both the frequency and time domains \cite{Taflove2005}. A simple example is the Drude-Lorentz function for a normal metal where, at low frequencies, the conductivity in the frequency domain has the form
\begin{equation}
\sigma(\omega)\equiv \sigma_{1}(\omega)+\rmi\sigma_{2}(\omega)=\frac{\sigma_0}{1-\rmi\omega\tau},\label{eq1}
\end{equation}
giving an absorptive part $\sigma_{1}(\omega)=\sigma_0\cdot 1/[1+(\omega\tau)^2]$ and reactive part $\sigma_{2}(\omega)=\sigma_0\cdot \omega\tau/[1+(\omega\tau)^2]$. In the time domain, the Drude-Lorentz response takes the form
\begin{equation}
\sigma(t)=\frac{\sigma_0}{\tau}\cdot\rme^{-t/\tau} \label{eq2}
\end{equation}
or, for the susceptibility
\begin{equation}
\chi(t) = \sigma_0\cdot\left(1-\rme^{-t/\tau}\right), \label{eq3}
\end{equation}
where $\sigma_0=\omega_p^2\tau\epsilon_0$ is the conductivity at zero frequency and $\omega_p=\sqrt{ne^2/m\epsilon_0}$ is the plasma frequency, yielding $\sigma_0=ne^2\tau/m$. Here $n$ is the density of charge carriers having charge $e$ and mass $m$. The scattering time $\tau$ is the mean time between collisions (dominated by elastic scattering in a so-called dirty metal). Note that we have not included a  Heaviside step function to ensure the convolution integration (for determining the response) only includes contributions for prior times.   

In the THz range, superconductors offer unique electromagnetic characteristics, such as a threshold for absorption and a nearly perfect inductive response at lower frequencies \cite{Tinkham2004}. In a conventional BCS-type superconductor, the absorption edge represents the energy threshold for a photon to break apart a Cooper pair, corresponding to a photon energy $\hbar\omega_g = 2\Delta$. If one is dealing strictly with frequencies substantially below the absorption threshold, then the response of a superconductor is well described by a Drude-Lorentz function where the mean scattering time approaches infinity. The result is a $\delta$-function form for $\sigma_1(\omega)$ and a corresponding $\sim 1/\omega$ behavior for $\sigma_2(\omega)$. While such a functional form is suitable for the low frequency spectral range, it will not suffice when the spectral range spans the energy gap. To avoid this deficiency, we have developed a model function, with an explicit dependence on the BCS gap, that approximates the response of a superconductor. Such a function could be used in the simulation of superconducting metamaterials \cite{Anlage2011}, especially those made from BCS superconductors \cite{Jin2010}. It could also be used to study nonlinear microwave effects in superconductors, as will be demonstrated in this paper.

\section{Development of model $\chi_s(t)$ for a superconductor}
\begin{figure}[t]
\centering
\includegraphics[scale=1]{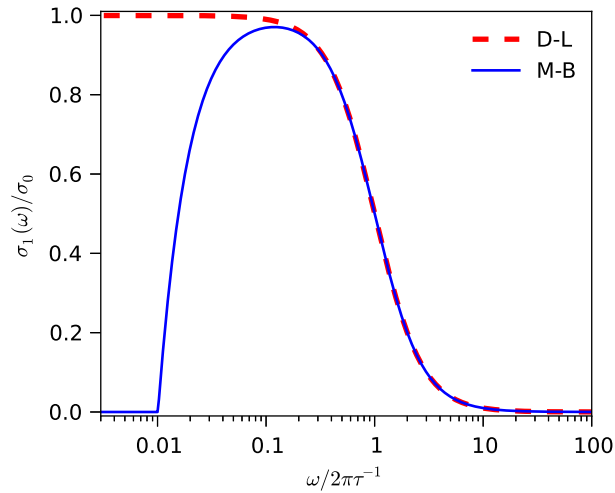}                
\caption{Real parts of the frequency dependent conductivity for the Drude-Lorentz function (\dashed) for the normal state and the Mattis-Bardeen theory (\full) for a BCS superconductor using $\omega_g/2\pi=1$~THz and $2\pi/\tau=100\omega_g$, both normalized to the DC conductivity $\sigma_0$.} 
\label{Fig1}
\end{figure}

As indicated above, our intent is to develop a convenient $\chi_s(t)$ expression containing the fundamental response characteristics for a superconductor, specifically, the pure inductive response of the superfluid condensate at low frequencies and an absorption gap at intermediate frequencies. Commonly used superconducting materials for electrodynamic applications are niobium and its compounds such as Nb:TiN and NbN, all of which are classified as BCS-type superconductors. Therefore, a useful starting point is the well-established, frequency-dependent conductivity given by Mattis and Bardeen \cite{Mattis1958}. Their expressions give both real and imaginary parts as ratios to the normal-state DC conductivity in the limit that the normal-state scattering rate $1/\tau$ is large compared to the energy gap frequency $\omega_{g}=2\Delta/\hbar$, the so-called ``dirty limit''. The real part is shown in \fref{Fig1}. The relevant normal-state Drude-Lorentz conductivity is shown for comparison. For frequencies $\omega\gg\omega_{g}$, the Mattis-Bardeen conductivity asymptotically approaches the Drude-Lorentz model, i.e $\sigma_1(\omega\gg\omega_g)\sim\sigma_0/[1+(\omega\tau)^2]$. Note that all of the physical information can be found in the real part since the imaginary part follows from a Kramers-Kronig transform. Since many applications have the superconductor's temperature $T$ well below its transition temperature $T_c$, we adopt the $T\rightarrow 0$ limit. Thus our approach does not include the contributions from broken pair (quasiparticle) excitations. These could be thermally excited quasiparticles, responsible for residual microwave dissipation as occurs in resonant cavities. Or they could be photo-excited quasiparticles created by an incident THz wave having spectral content greater than the optical gap frequency. Therefore, our model loses validity when either of these quasiparticle contributions become significant.

\begin{figure}[t]
\includegraphics[scale=0.97]{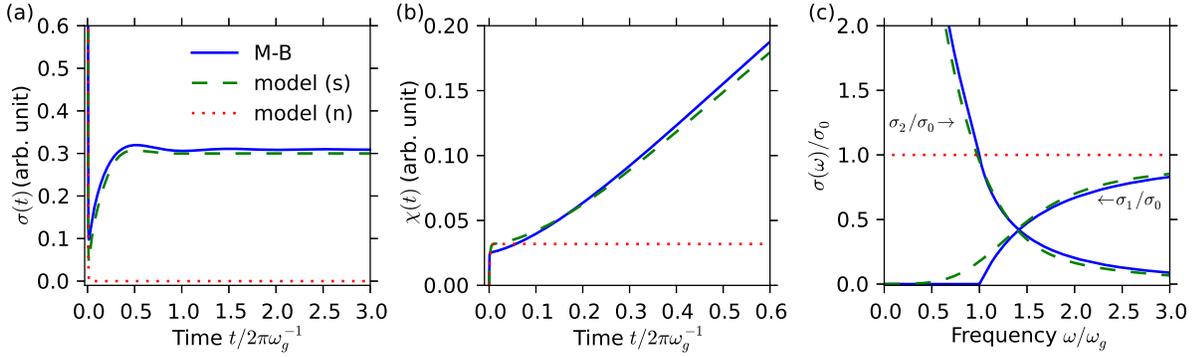}                
\caption{Comparisons between the Mattis-Bardeen (\full) and our approximate model response function (\dashed) with $\omega_g/2\pi=1$~THz and $\gamma=100\omega_g$. Also shown is the normal state response (\dotted). (a) The time-dependent conductivity. (b) The corresponding time-dependent susceptibility. (c) The frequency-dependent conductivity. Note that the results in (a) and (c) are connected through a Fourier transform.} 
\label{Fig2}
\end{figure}

A Fourier transform of the Mattis-Bardeen $\sigma(\omega)$ yields the corresponding $\sigma(t)$ shown in \fref{Fig2}(a). Note that the $\delta$-function response of the superfluid condensate must be included to maintain causality. The result is a $\sigma(t)$ that saturates to a finite positive quantity in the limit of very long times. Since the current is a convolution of the conductivity and E-field, a constant conductivity results in a current directly proportional to the time-integral of the E-field, as expected for a perfect inductor. Also, the strength of the superfluid condensate is determined from the ``missing area'' in the real conductivity that develops when
the gap opens up. This missing area scales with the size of the energy gap as well as the normal-state DC conductivity, thus this positive constant in $\sigma(t)$ at $t\gg 2\pi/\omega_g$ should have factors of both $\omega_g$ and $\sigma_0$. This form for $\sigma(t)$ at long times $t$ can also be understood from the viewpoint of a conventional metal with an infinitely long time $\tau$ between scattering events. Expanding the Drude-Lorentz conductivity of equation~\eref{eq2} in a power series
\begin{equation}
\sigma(t)= \frac{\sigma_0}{\tau}\cdot e^{-t/\tau} = \frac{ne^2}{m}\cdot\left[1-t/\tau+\frac{1}{2}\cdot (t/\tau)^2+\ldots\right]
\end{equation}
and taking the limit of $\tau\rightarrow \infty$ yields $\sigma(t)=ne^2/m=$~constant. 

Integrating the $\sigma(t)$ in \fref{Fig2}(a) yields the time-domain susceptibility in \fref{Fig2}(b). On short time scales one observes the susceptibility following the Drude-Lorentz form $1-\rme^{-t/\tau}$ as expected. Then, for intermediate times, the susceptibility changes over to linear in time (with zero intercept), including some damped oscillations (more apparent in $\sigma(t)$ than $\chi(t)$) at the gap period $T_g=2\pi/\omega_g$ and related to the opening of the gap in $\sigma_1(\omega)$. Thus, for our approximate model response, we seek a function for $\chi(t)$ that behaves as $1-\rme^{-t/\tau}$ for short times and $A\cdot t$ (where $A$ is proportional to $\omega_g$) for $t\gg 2\pi/\omega_g$, with the changeover occurring in the time range corresponding to $t\sim 2\pi/\omega_g$. A simple functional form having most of these characteristics can be written as
\begin{equation}
\chi_s(t) = \sigma_0\cdot \left[C_1\cos(k_1 t)\mathrm{e}^{-k_2t} + C_2\mathrm{e}^{-k_3t} - \mathrm{e}^{-t/\tau}+C_3k_4t\right],\label{eq4}
\end{equation}
where a good fit to the Mattis-Bardeen conductivity (shown in \fref{Fig2}(a) and \fref{Fig2}(b)) is achieved with $k_1=k_2=k_4=\omega_g$, $k_3=4\omega_g$ along with $C_1=\frac{5}{6}$, $C_2=\frac{1}{6}$ and $C_3=\frac{3}{2}$. Note that the $C_1$ and $C_2$ terms form a gap in the Drude-Lorentz response while the $C_3$ term gives the superfluid response. With $\omega_g$ set to zero, the normal-state Drude-Lorentz response of equation~\eref{eq3} is recovered.  

When differentiated to yield the conductivity, one finds that this functional form allows for an explicit Fourier transform, yielding a closed-form expression in the frequency domain. By comparing to the same functional form with $\omega_g=0$, one can integrate the missing area in $\sigma_1(\omega)$ to verify that $C_3=\frac{3}{2}$ and $k_4=\omega_g$ are consistent with the Ferrell-Glover-Tinkham sum rule \cite{Ferrell1958}. We therefore have the following approximate time-domain susceptibility for a weak-coupled BCS superconductor and $T\ll T_c$:
\begin{equation}
\chi_s(t) = \sigma_0\cdot \left[\frac{5}{6}\cos(\omega_g t)\mathrm{e}^{-\omega_gt} + \frac{1}{6}\mathrm{e}^{-4\omega_gt} - \mathrm{e}^{-t/\tau}+\frac{3}{2}\omega_gt\right].\label{eq5}
\end{equation}
The time derivative of this susceptibility yields the time-domain conductivity, which has an equivalent expression in the frequency domain, namely:
\begin{eqnarray}
\fl\frac{\sigma_{1s}(\omega)}{\sigma_0} = -\left(\frac{5}{6}\right)\frac{1}{1+\frac{1}{4}(\frac{\omega}{\omega_g})^4}-\left(\frac{1}{6}\right)\frac{1}{1+\frac{1}{16}(\frac{\omega}{\omega_g})^2}+\frac{1}{1+\omega^2\tau^2}+\frac{3}{2}\omega_g\delta(0), \label{eq6}\\
\fl\frac{\sigma_{2s}(\omega)}{\sigma_0} =  -\left(\frac{5}{6}\right)\frac{\frac{1}{2}\frac{\omega}{\omega_g}+\frac{1}{4}(\frac{\omega}{\omega_g})^3}{1+\frac{1}{4}(\frac{\omega}{\omega_g})^4} -\left(\frac{1}{6}\right)\frac{\frac{1}{4}\frac{\omega}{\omega_g}}{1+\frac{1}{16}(\frac{\omega}{\omega_g})^2}-\frac{\omega\tau}{1+\omega^2\tau^2}+\frac{3}{2}\frac{\omega_g}{\omega}.\label{eq7}
\end{eqnarray}
A comparison of these expressions with the exact Mattis-Bardeen conductivity is given in \fref{Fig2}(c). Unlike the abrupt absorption edge in the Mattis-Bardeen conductivity, the approximate model conductivity of equation~\eref{eq6} has a more gradual increase at the gap frequency along with a small ``tail'' extending to $\omega=0$. Above the gap frequency, the model agrees well with the Mattis-Bardeen theory. The extra absorption near the gap frequency and weak tail to low frequencies cause the model to underestimate slightly the $\delta$-function strength when compared with the Mattis-Bardeen theory. 

\section{Application examples using the model $\chi_s(t)$}
The frequency-domain expression of the model conductivity or susceptibility can be conveniently used for approximating the optical response of a BCS-type superconductor when the exact integral expressions are too computationally intensive. However, we expect the time-domain conductivity or susceptibility functions to be even more useful in FDTD calculations. We demonstrate this in the following examples. 

Consider an incident electromagnetic wave propagating and interacting with a superconductor. For simplicity we look at the problem in one dimension, but the method is also applicable in 2-D and 3-D. The wave propagation can be simulated using the FDTD method \cite{Yee1966,Taflove2005} that iteratively solves the Maxwell curl equations over space and time,
\begin{eqnarray}
\nabla\times H = \frac{\partial D}{\partial t}, \label{eq8}\\
\nabla\times E = -\frac{\partial B}{\partial t}, \label{eq9}
\end{eqnarray}
where the step from $D(t)$ to $E(t)$, assuming a linear response, involves the convolution, 
\begin{equation}
D(t) = \epsilon_{\infty}\epsilon_0 E(t) +\epsilon_0\int_0^{t} E(t-\tau)\chi(\tau)\rmd \tau, \label{eq10}
\end{equation}
with $\epsilon_{\infty}$ being the high-frequency dielectric constant, $\chi(t)$ the superconductor susceptibility given by equation~\eref{eq5}, and all quantities are for a given spatial coordinate. Discretizing equation~\eref{eq10} for FDTD calculations in the same manner as in reference \cite{Luebbers1990} has $t$ substituted with $k\cdot \Delta t$ where $\Delta t$ is the FDTD step size (and not the product of the BCS gap and time) and $k$ is the step index. Using the notation $D(k\cdot\Delta t)\equiv D^k$, we have
\begin{eqnarray}
    D^k &= \epsilon_{\infty}\epsilon_0 E^k +\epsilon_0\sum_{m=0}^{k-1}E^{k-m}\int_{m\Delta t}^{(m+1)\Delta t}\chi(\tau)\rmd \tau \nonumber\\
        &\equiv \epsilon_{\infty}\epsilon_0 E^k+ \epsilon_0\sum_{m=0}^{k-1}E^{k-m}\chi^m, \label{eq11} 
\end{eqnarray}
from which $E^k$ can be solved from $D^k$.

The above method is inefficient because it requires using $E^k$ and calculating $\chi^k$ for all previous time steps. To improve calculation efficiency, a recursive convolution approach \cite{Luebbers1990,Taflove2005} can be used that takes advantage of the simple form of the susceptibility function~\eref{eq5}, yielding
\begin{equation}
D^k = \epsilon_0 \left(\epsilon_{\infty} + \sum_{j=1}^5 \chi_j^0 \right)E^k + \sum_{j=1}^5 \rme ^{\gamma_j\Delta t}\phi_j^{k-1}+\frac{3}{2}\epsilon_0\sigma_0\omega_g(\Delta t)^2\eta^{k-1}, \label{eq12}
\end{equation}
where $\gamma_j$ and $\chi_j^m$ are defined in \tref{table1}, $\phi_j^k=\epsilon_0\sum_{m=0}^{k-1}E^{k-m}\chi_j^m$ and $\eta^k=\sum_{m=1}^k E^m$.
\Eref{eq12} only involves two quantities from the previous time step, $\phi_j^{k-1}$ and $\eta^{k-1}$. This significantly reduces the amount of calculation required by equation~\eref{eq11}. At the time step $k$, $\eta^k$ is updated as $\eta^k = \eta^{k-1}+E^k$ and $\phi_j^k$ updated using the recursive relations listed in \tref{table1}.

\begin{table}
\caption{\label{table1}Coefficients, terms and recursive relations used in the recursive convolution.}
\begin{indented}
\item[]\begin{tabular}{@{}llll}
\br
$j$ & $\gamma_j$ &  $\chi_j^m$ & Recursive relation for $\phi_j^k$ \\
\mr
1& $(\rmi-1)\omega_g$ & $[5(1+\rmi)\sigma_0/24\omega_g](1-\rme^{\gamma_1\Delta t})\rme^{m\gamma_1\Delta t}$ & \\
2& $-(\rmi+1)\omega_g$& $[5(1-\rmi)\sigma_0/24\omega_g](1-\rme^{\gamma_2\Delta t})\rme^{m\gamma_2\Delta t}$ & $\phi_j^k=\epsilon_0\chi_j^0E^k+\rme^{\gamma_j\Delta t}\phi_j^{k-1}$\\
3& $-4\omega_g$ & $(\sigma_0/24\omega_g)(1-\rme^{\gamma_3\Delta t})\rme^{m\gamma_3\Delta t}$ & $(j=1,2,3,4)$ \\
4& $-1/\tau$ &  $-\sigma_0\tau (1-\rme^{\gamma_4\Delta t})\rme^{m\gamma_4\Delta t}$ & \\
5& 0 & $\frac{3}{4}(2m+1)\sigma_0\omega_g(\Delta t)^2$ & $\phi_5^k = \epsilon_0\chi_5^0E^k + \phi_5^{k-1}$\\
&&& $\qquad +\frac{3}{2}\epsilon_0\sigma_0\omega_g(\Delta t)^2\eta^{k-1}$\\
\br
\end{tabular}
\end{indented}
\end{table}

\begin{figure}[b]
\includegraphics[scale=1]{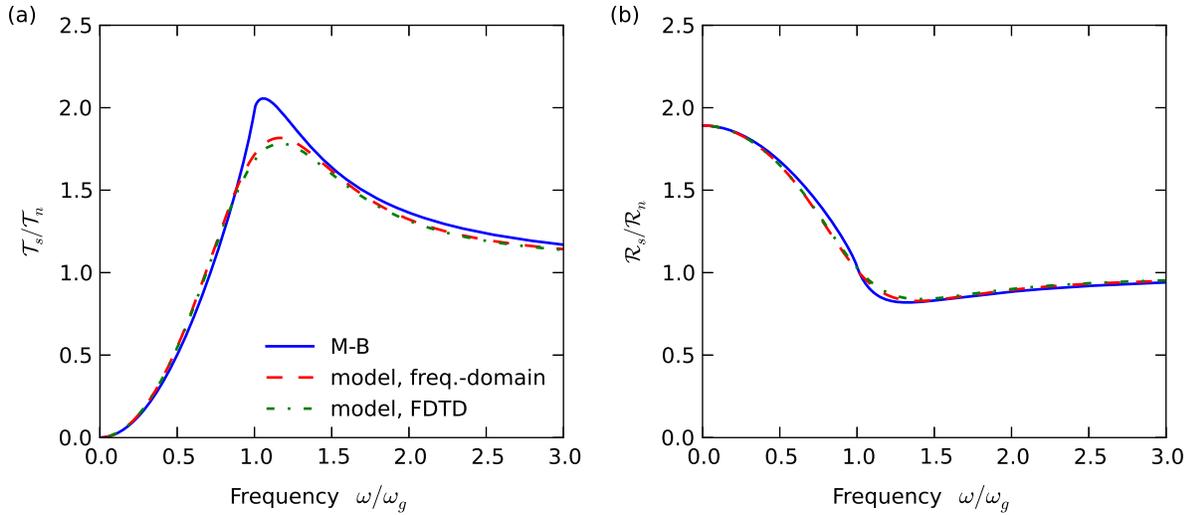}                
\caption{Superconducting-to-normal transmission (a) and reflection (b) ratios for a thin-film superconductor with $\omega_g/2\pi=1$~THz and $\gamma=100\omega_g$ on a dielectric substrate. Results are calculated using the Mattis-Bardeen conductivity (\full), the model frequency-domain conductivity function (\dashed), and by FDTD method using the model time-domain susceptibility function (\chain).} 
\label{Fig3}
\end{figure}  

\subsection{Transmission and reflection for a thin film}
\label{sec_TRlinear}
To illustrate the capabilities of this simple model, we use it in an FDTD calculation for the transmitted and reflected electromagnetic waves for a thin-film normal metal and a thin-film superconductor on a dielectric substrate. For simplicity, we use an incident single-cycle waveform having spectral content spanning the superconductor's energy gap, and do not include the effects of multiple internal reflections inside the substrate. The calculation is performed on a 1-D grid using the recursive convolution approach described above, with the film occupying a single grid position, consistent with the film being thinner than any relevant length scale. The parameters used are $\omega_g/2\pi=1$~THz, film sheet resistance 115~$\Omega/\square$, and the substrate refractive index 3.05. A grid spacing $\le$1~$\mu$m yields consistent result (0.5~$\mu$m is used in the example here). The calculation yields both transmitted and reflected waveforms, from which we determine the reflection from the front surface as well as the transmission into the substrate  for both superconducting and normal states. The superconducting-to-normal state transmission and reflection ratios are shown as the green dash-dot curves in \fref{Fig3}.  

We also calculate the same quantities using the frequency-domain conductivity equations \eref{eq6} and \eref{eq7} as well as using the actual Mattis-Bardeen conductivity with the well-known thin-film transmission and reflection expressions given by Glover and Tinkham \cite{Glover1957}. These are also plotted in \fref{Fig3} for comparison with the FDTD results. Agreement is good, except for very near the energy gap frequency where our model function has a more gradual onset of absorption. Indeed, experimental results often suggest a slower absorption onset, possibly a result of anisotropy or inhomogeneity.

\subsection{Bulk reflectance}
\label{sec_bulkR}
\begin{figure}[t]
\includegraphics[scale=1]{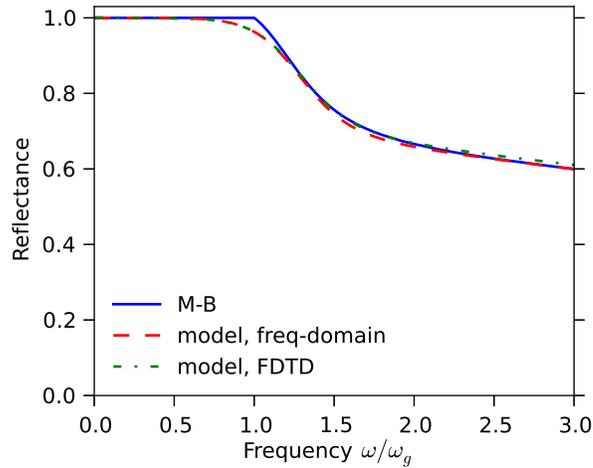}  
\centering              
\caption{Reflectance of a bulk superconductor with $\omega_g/2\pi=1$~THz and $\omega_p=\gamma=100\omega_g$ calculated using the Mattis-Bardeen conductivity (\full), the model frequency-domain conductivity function (\broken), and by FDTD method using the model time-domain susceptibility function (\chain).} 
\label{Fig4}
\end{figure}

The thin film calculation discussed above can be used to calculate the reflectance of a bulk superconductor. Here, the superconductor necessarily occupies multiple grid locations with a grid spacing significantly smaller than, and to a depth much greater than, the penetration depth. One can determine both by reducing the grid spacing and increasing the grid depth until the reflected waveform no longer changes. The FDTD result, using a 0.5~$\mu$m grid spacing and 100 grid points into the superconducting material, is shown in \fref{Fig4}. For comparison, we also calculate the bulk reflectance using the Mattis-Bardeen theory and the frequency-domain conductivity equations~\eref{eq6} and~\eref{eq7}. The model calculation results are in good overall agreement with the Mattis-Bardeen theory. Again, the main discrepancy occurs near the gap frequency where the expected sharp kink is rounded due to the absence of an abrupt absorption edge in the model.

\section{Strong fields and nonlinear microwave response}
The electromagnetic response of a superconductor can become nonlinear when the incident field is sufficiently strong \cite{Samoilova1995} or when resonances occur such as for plasmonic structures. For the spectral range $\omega>\omega_g$, absorption occurs as Cooper pairs are broken. For $\omega<\omega_g$, although
the photons cannot directly break pairs, the E-field drives supercurrents. Though only the former involves absorption and dissipation, both lead to a weakened superconducting state \cite{Owen1972,PB} and smaller energy gap $\Delta$. Since the susceptibility is a function of $\Delta$, a nonlinear response will occur for both cases if the light-induced perturbation is significant. Here we focus on the case when the spectral content of the incident electromagnetic wave has $\omega<\omega_g$, e.g. at microwave frequencies where supercurrents dominate the response. Thus our model for non-linear response is not valid for an intense source having spectral content spanning the gap frequency and resulting in direct pair-breaking to yield a large quasiparticle density. We note that 3rd harmonic generation in BCS superconductors has been observed using microwave techniques and theoretically shown to result when no biasing current or magnetic field is present \cite{Samoilova1995}. Here we show how our time-domain susceptibility model can be used to calculate the response and the nonlinear production of odd harmonics. The calculation is an extension of the method in Section~\ref{sec_TRlinear} where we used the FDTD method to determine the transmitted E-field waveform through a thin-film superconductor. The material parameters are the same as used in Section~\ref{sec_TRlinear}. The difference is that we now allow the gap to vary as a function of the current density $J(t)$ induced in the superconductor, i.e. $\chi_s[t,\Delta(J)]$. We begin by considering a model where the energy gap varies according to Ginzburg-Landau theory and the Boltzmann equation \cite{Geier1982}, i.e.
\begin{equation}
\frac{\partial f}{\partial t}=\frac{1}{2\tau_{\Delta}}\left(1-f^2-\frac{j^2}{f^4}\right).\label{eq_nl}
\end{equation}
where $f=\omega_g(t)/\omega_g(0)$ (with $\omega_g(0)$ the unperturbed gap), $j=\sqrt{4/27}J/J_c$ (with $J$ the field-induced current density and $J_c$ the critical current density), and $\tau_{\Delta}=3.7(kT_c/\Delta)\tau_E$ is the order parameter relaxation time (with $\Delta=\hbar\omega_g(0)/2$ and $\tau_E$ the electron-phonon scattering time). Using $T_c=13$~K, $\omega_g(0)/2\pi=1$~THz and $\tau_E=10$~ps \cite{Ilin2000} for NbN, we estimate $\tau_{\Delta}\approx 20$~ps. The current density $J$ is calculated as the convolution of the E-field with the conductivity $\sigma(t)$, i.e. the time derivative of equation~\eref{eq5}.
 
\begin{figure}[t]
\includegraphics[scale=0.95]{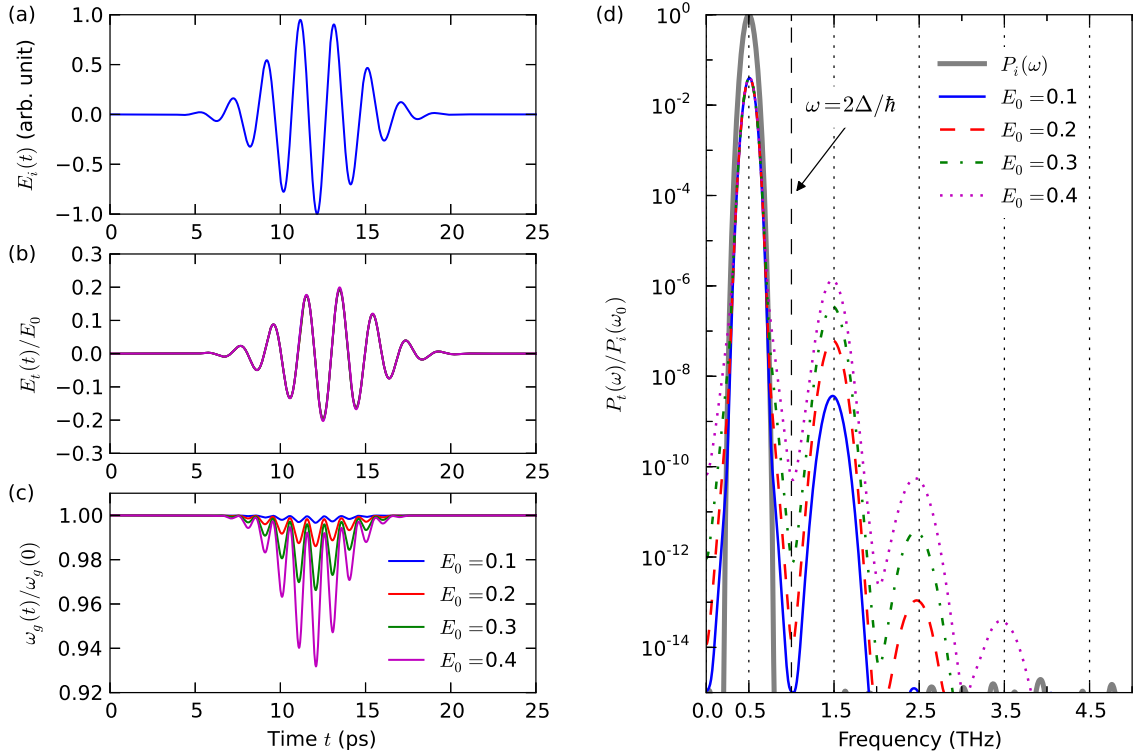}  
\centering              
\caption{(a) Time-domain waveform of the incident field $E_i(t)$. (b) Time-domain transmitted waveforms $E_t(t)$ through a superconducting thin film, normalized to the incident wave amplitude $E_0$. (c) Time dependence of the gap $\omega_g(t)$, normalized to the unperturbed gap $\omega_g(0)$, associated with the waveforms in (b). (d) Power spectra of the incident $P_i(\omega)$ and the transmitted $P_t(\omega)$ waveforms for different $E_0$ values, all normalized to $P_i(\omega_0)$ where $\omega_0=0.5$~THz.}
\label{Fig5}
\end{figure}  
 
The incident time-domain waveform for the FDTD calculation is chosen to be a several-cycle pulse with center frequency at 0.5~THz, as shown in \fref{Fig5}(a). The spectral intensity of this incident waveform is shown as the solid gray curve in \fref{Fig5}(d). Since we intend a Fourier analysis of the resulting waveform's spectral content, we chose a waveform envelope given by the 7-term Blackman-Harris function commonly used for apodization in Fourier spectroscopy to avoid finite window artifacts. The pulse was made incident onto the superconducting film and the induced current density calculated for each time step. As described above, the gap frequency and, from that, the susceptibility $\chi(t)$ were updated for the next time step according to equations \eref{eq_nl} and \eref{eq5}. The amplitude $E_0$ of the incident wave was chosen such that $E_0 = 1.0$ yielded a maximum induced current density equal to $J_c$ for NbN.  We then calculated the transmitted waveforms for incident amplitudes $E_0$ varying from 0.1 to 0.4 such that the critical current density was never reached. Figure~\ref{Fig5}(b) shows the resulting transmitted waveforms $E_t(t)$ through the film, each normalized to the incident $E_0$. Note that the nonlinear behavior is weak on the scale of $E_0$. Figure~\ref{Fig5}(c) shows the time dependence of the gap $\omega_g(t)$, normalized to $\omega_g(0)$, as the incident wave propagates through the film. As expected, the model yields an energy gap suppressed by the induced current, becoming stronger as the incident field amplitude increases. The resulting nonlinear effect can be seen in the transmitted wave's power spectrum $P_t(\omega)\equiv |E_t(\omega)|^2$, shown in figure~\ref{Fig5}(d). On this log scale, $P_t(\omega)$ shows clearly the 3rd harmonic for all four $E_0$ values. The higher odd harmonics become stronger with increasing $E_0$. Though not shown, the degree of nonlinear upconversion also varies with $\tau_{\Delta}$, providing sensitivity to quantities such as the inelastic scattering time for electrons. We remind the reader that the integral expression connecting the susceptibility and electric field to yield the polarization (equation~\eref{eq10}) is no longer exact when the susceptibility is nonlinear \cite{Boyd2008}. Our example calculation corresponds to keeping only the leading term in a convolution series, which is probably sufficient for the small degree of nonlinear upconversion that resulted.
 
Though we have done our calculation for the transmission through a thin film, the model is suitable for determining the nonlinear response when the incident waveform is reflected from a bulk superconductor as in Section~\ref{sec_bulkR}. In that case, the current density in each discretized layer into the superconductor's surface would be used to determine the energy gap change for that layer, as described above. With the superconducting coherence length smaller than the layer thickness, each layer can have a distinct gap and therefore a distinct local response.

\section{Conclusion}
We have developed a simple approximate function for the time-domain susceptibility of a dirty-limit BCS superconductor, appropriate for the spectral range both below the optical gap frequency (where the superfluid response dominates) and above the gap frequency (where it asymptotes to the Drude response).  As demonstrated, the functional form can be used in FDTD calculations using the recursive convolution approach. The expression does not include the contribution from thermal quasiparticles nor from additional quasiparticles generated by the incident wave through direct pair breaking. Therefore, the expression's validity is limited to $T\ll T_c$ and a linear response when the spectral content exceeds the gap frequency.

If the spectral range is restricted to $\omega < \omega_g$, e.g. microwaves, it allows FDTD calculations for the nonlinear response when strong induced currents weaken the superconducting state. Ginzburg-Landau theory is used to relate the current to a change in the superconducting gap, 
including an order parameter (gap) relaxation time.

More complicated functional forms that improve the absorption onset, manage the intermediate and clean-limits, or include absorption by thermal quasiparticles (broken pairs) can probably be developed, but the features of our simple functional form should be sufficient for many applications.

\ack
This work was supported by the U.S. Department of Energy through contract DE-AC02-98CH10886 at BNL.


\section*{References}
\begin{thebibliography}{99}
\bibitem{Lal2007} Lal S, Link S and Halas N J 2007 {\it Nat. Photon.} {\bf 1} 641; Feigenbaum E and Orenstein M  2007 {\it J. Lightwave Technol.} {\bf 25} 2547; Feise M W, Schneider J B and Bevelacqua P J 2004 {\it IEEE Trans. Antennas Propagat.} {\bf 52} 2955; Ahmadi A and Mosallaei H 2008 {\it \PR}B {\bf 77} 045104; Ziolkowski R W and Heyman E 2001 {\it \PR}E {\bf 64} 056625

\bibitem{Semouchkina2005} Semouchkina E A, Semouchkin G B, Lanagan M and Randall C A 2005 {\it IEEE Trans. Microwave Theory Tech.} {\bf 53} 1477; Wongkasem N, Akyurtlu A, Li J, Tibolt A, Kang Z and Goodhue W D 2006 {\it Prog. Electromagn. Res. PIER} {\bf 64} 205; Hosseini A, Nejati H and Massoud Y 2008 {\it Opt. Express} {\bf 16} 1475; Yanai A and Levy U 2009 {\it Opt. Express} {\bf 17} 924; Dorfm\"{u}ller J, Vogelgesang R, Khunsin W, Rockstuhl C, Etrich C and Kern K 2010 {\it Nano Lett.} {\bf 10} 3596 

\bibitem{Taflove2005} Taflove A and Hagness S C 2005 {\it Computational Electrodynamics: the Finite-Difference Time-Domain Method} (Norwood, MA: Artech House) Chap.~9  

\bibitem{Tinkham2004} Tinkham M 2004 {\it Introduction to Superconductivity} (Mineola, NY: Dover Publications)

\bibitem{Anlage2011} Anlage S M 2011 {\it J. Opt.} {\bf 13} 024001

\bibitem{Jin2010} Jin B B, Zhang C H, Engelbrecht S, Pimenov A, Wu J B, Xu Q Y, Cao C H, Chen J, Xu W W, Kang L and Wu P H 2010 {\it Opt. Express} {\bf 18} 17504; Zhang C H, Wu J B, Jin B B, Ji Z M, Kang L, Xu W W, Chen J, Tonouchi M and Wu P H 2012 {\it Opt. Express} {\bf 20} 42  

\bibitem{Mattis1958} Mattis D C and Bardeen J 1958 {\it \PR}{\bf 111} 412; see also Zimmermann W, Brandt E H, Bauer M, Seider E and Genzel L 1991 {\it Physica} C {\bf 183} 99 and Leplae L 1983 {\it \PR}B {\bf 27} 1911 for the extension to arbitrary scattering rates  

\bibitem{Ferrell1958} Ferrell R A and Glover R E, III 1958 {\it \PR}{\bf 109} 1398; Tinkham M and Ferrell R A 1959 {\it \PRL}{\bf 2} 331 

\bibitem{Yee1966} Yee K S 1966 {\it IEEE Trans. Antennas Propag.} {\bf 14} 302 

\bibitem{Luebbers1990} Luebbers R, Hunsberger F P, Kunz K S, Standler R B and Schneider M 1990 {\it IEEE Trans. Electromagn. Compat.} {\bf 32}  222; Beard M C and Schmuttenmaer C A 2001 {\it J. Chem. Phys.} {\bf 114} 2903

\bibitem{Glover1957} Glover R E, III and Tinkham M 1957 {\it \PR}{\bf 108} 243 

\bibitem{Samoilova1995} Samoilova T B 1995 {\it Supercond. Sci. Technol.} {\bf 8} 259

\bibitem{Owen1972} Owen C S and Scalapino D J 1972 {\it \PRL} {\bf 28} 1559; Carr G L, Lobo R P S M, LaVeigne J D, Reitze D H and Tanner D B 2000 {\it \PRL} {\bf 85} 3001

\bibitem{PB} Maki K 1969 {\it Gapless Superconductivity} ({\it Superconductivity} vol 2) ed Parks R D (New York, NY: Marcel Dekker); Skalski S, Betbeder-Matibet O and Weiss P R 1964 {\it Phys. Rev.} {\bf 136} A1500; Anthore A, Pothier H and Esteve D 2003 {\it Phys. Rev. Lett.} {\bf 90} 127001; Xi X, Hwang J, Martin C, Tanner D B and Carr G L 2010 {\it Phys. Rev. Lett.} {\bf 105} 257006

\bibitem{Geier1982} Geier A and Sch\"{o}n G 1982 {\it J. Low Temp. Phys.} {\bf 46} 151

\bibitem{Ilin2000} Il'in K S, Lindgren M, Currie M, Semenov A D, Gol'tsman G N, Sobolewski R, Cherednichenko S I and Gershenzon E M 2000 {\it Appl. Phys. Lett.} {\bf 76} 2752

\bibitem{Boyd2008} Boyd R W 2008 {\it Nonlinear Optics} (Burlington, MA: Academic Press) Chap. 1

\end{thebibliography}
\end{document}